\documentstyle [12pt,aasms4] {article}

\newcommand{\msun}{~{\rm M}_\odot}
\def\lsim{\mathrel{\rlap{\lower4pt\hbox{\hskip1pt$\sim$}}
	\raise1pt\hbox{$<$}}}		
\def\gsim{\mathrel{\rlap{\lower4pt\hbox{\hskip1pt$\sim$}}
	\raise1pt\hbox{$>$}}}		

\begin{document}

\title{ EVOLUTION OF BINARY COMPACT OBJECTS WHICH MERGE }
\author{ HANS A. BETHE }
\affil{Floyd R. Newman Laboratory of Nuclear Studies \\
	Cornell University \\
	Ithaca, New York 14853, USA}
\author{ G. E. BROWN}
\affil{Department of Physics and Astronomy \\
	State University of New York\\
	Stony Brook, New York 11794-3800, USA}

\begin{abstract}
{ Beginning from massive binaries in the Galaxy we evolve black--hole,
neutron--star binaries and binary neutron stars, such as the Hulse--Taylor
system PSR 1913+16.  The new point in our evolution is a quantitative
calculation of the accretion of matter by a neutron star in common envelope
evolution which sends it into a black hole.  We calculate the mass of the latter
to be $\sim 2.4\msun$.

The black hole fate of the first neutron star can only be avoided if the
neutron star does not go through common envelope evolution.  This can be
realized if the two massive binaries are sufficiently close in mass, not more
than $\sim$ 5\% apart, so that they burn helium at the same time.  Then their
common hydrogen envelope is expelled by the tightening in the double He--star
system, with attendant hydrodynamical coupling to the envelope.  In this way
we obtain a rate of $\lsim 10^{-5}$ per year per Galaxy for production of the
narrow neutron--star binaries such as the Hulse--Taylor 1913+16 or the
Wolszczan 1934+12.  This is in agreement with estimates based on the observed
number of such systems extrapolated to the entire Galaxy with beaming factors
and corrections for the $\sim$ 90\% of the binary pulsars estimated to be
unobservable.

Our chief conclusion is that the production rate for black--hole,
neutron--star binaries (in which the neutron star is unrecycled) is $\sim
10^{-4}$ per year per Galaxy, an order of magnitude greater than that of
neutron star binaries.  Not only should this result in a factor of $\sim$ 10
more mergings for gravitational wave detectors like LIGO, but the signal should
be larger.

We give some discussion of why black--hole, neutron--star binaries have not
been observed, but conclude that they should be actively searched for.
}
\end{abstract}
\keywords{black holes -- compact objects -- mergers
-- gravitational waves}

\section{ INTRODUCTION}

LIGO, an elaborate system of detectors, is being prepared to measure
gravitational waves.  It is generally believed that the best source of
such waves is the merger of two compact stars, neutron stars (NS) or
black holes (BH).  It is therefore useful to predict the frequency of
such mergers.

The current belief is that there are about $10^{-5}$ mergers per
Galaxy per year.  We wish to show that the rate is about 10 times
greater, of order $10^{-4}$ per Galaxy per year.

We wish to show that a substantial fraction of supernovae in binary
stars will head in time to two compact stars which will ultimately
merge.  Our assumptions are listed in \S 2.

In \S 3 we estimate the fraction of binaries which contain two stars
massive enough to become supernovae provided they contain one.  We
find this fraction to be about 50\%.

The most likely way in which a binary of interest may be disrupted is
by the kick which the NS receives in the two SN (supernovae).  We use
recent observations by Chernoff and Cordes (1997) of pulsar
velocities, and find that about 40\% of binaries survive that kick (\S
2).

We find (\S 4) that in 80\% of the binaries of interest, the first SN,
in star $A$, occurs while the other star ($B$) is still in its main
sequence (MS).  Subsequently star $B$ evolves into a giant.  The NS,
$A$, will accrete part of the envelope of $B$ and thereby becomes a
BH.  The rest of the envelope of $B$ will be ejected, the energy for
this being provided by $A$ spiralling in (\S 4).  We require that $A$
stay outside the He core of star $B$, so that we ultimately get two
separate compact stars.  This depends importantly on $A$ becoming a
BH.

In the 20\% of cases where $B$ is already a giant when $A$ becomes a
SN, the evolution is somewhat more complicated (\S 6).  In the fraction of
these cases where stars $A$ and $B$ burn helium at the same time, the first
neutron star can escape common envelope evolution and so avoid going into a
black hole.  In these cases, a neutron--star binary such as the Hulse--Taylor
1913+16 can be formed, with total production probability for narrow
neutron--star binaries which merge of $\sim 10^{-5}$ per year per Galaxy.

In \S 7, we determine the maximum distance $R$ between the two compact
stars which will permit merger by gravitational wave emission within
a Hubble time.  Essentially all cases we discuss will permit such
merger.

We conclude (\S 8) that $\sim$  1\% of all SN in binary stars will
lead ultimately to mergers; thus we predict $10^{-4}$ mergers per
Galaxy per year.

\section{ASSUMPTIONS}

We call the star which is initially heavier, star $A$, the other star
$B$.  We denote initial masses by subscript $O$, so we have masses
$M_{AO}$, $M_{BO}$.  We denote their ratio by $q$, thus
$$
q = M_{BO}/M_{AO} \leq 1~~.
\eqno(2.1)
$$
\noindent Following Portegies Zwart and Yungelson (1998), we assume that $q$ is
distributed uniformly between 0 and 1.  Likewise following them, we
assume that log $a$ is uniformly distributed where $a$ is
the semi-major axis of their orbit.

However, we assume different limits for $a$ than they.
Initially both stars are massive main sequence stars, radius at least
3 ${\rm R}_\odot$, so $a>6 {\rm R}_\odot = 4 \times 10^6$ km.  At the other end,
we require $a< 4 \times 10^9$ km.  This corresponds to an orbital
velocity of 30 km $s^{-1}$, the same as the earth around the sun, and
an orbital period of 25 years.  At a smaller orbital velocity (longer
period), it would be difficult to recognize the stars as a binary.
Then the fraction of binaries in a given interval of $\ell n~a$ is
$$
d \phi = d (\ell n~a)/7~~.
\eqno(2.2)
$$

In fact, our lower limit is too low since two ZAMS early $B$ stars of $8\msun$
each would fill their Roche lobes in an orbit of $9 {\rm R}_\odot$ rather than
$6 {\rm R}_\odot$.  Thus $10^7$ km would be more reasonable.  Furthermore, as
Eggleton (1998) also pointed out, our flat distribution over log P is not well
supported for massive stars.  Garmany, Conti \& Massey (1980) did probably the
best analysis so far and found a binarity of 36 $\pm$ 7\% O--stars in a
magnitude--limited and declination--limited sample of 67 O--stars.  Longer
periods are largely inaccessible to measurement because of the inherent
instability of O--stars.  Although the observed binarity is smaller than our
assumed 50\%, the logarithmic interval in $a$ is also smaller so that the
number of binaries in a given logarithmic interval is not much changed.
Unfortunately, the interval $a_i$ which we find useful in eq. (5.24) presumably
includes some of the inaccessible binaries.  In the work referred to there is a
clustering of observed binaries in the period interval $\sim 3-10$ days, as
compared with $\sim 10-100$ days which we show below to be favorable for
gravitational merging with avoidance of in--spiral.  However, orbits widen
during mass transfer (RLOF) as we show below in our eq. (6.1), and the shorter
period stars will mostly widen into our favorable interval.  Thus, we believe
our flat distribution over log P with chosen interval and binarity to somewhat
underestimate the number of favorable binaries.

We assume that a star needs an initial mass
$$
M > M_S = 10 {\rm M}_\odot~~,
\eqno(2.3)
$$
\noindent to become a supernova.  Beyond this mass, we assume that the
rate of birth of stars of mass $M$ is proportional to
$$
\beta \sim M^{-n}~~,
\eqno(2.4)
$$
\noindent where $n$ is the Salpeter exponent, about 1.5.  The rate of
supernovae (SN) is the same as the rate of birth.  So if $\alpha$ is
the total rate of SN, the rate of SN in a mass interval $dM$ is
$$
d \alpha = \alpha n \left( M/10{\rm M}_\odot \right)^{-n} dM/M~~.
\eqno(2.5)
$$

We assume the total rate of supernova (SN) events per Galaxy to be
$$
\alpha^\prime = 2~ {\mbox{per century}} = 2 \times 10^{-2}~
{\rm{yr}}^{-1}~~.
\eqno(2.6)
$$
\noindent We assume that half\footnote{With our smaller interval in $a$ we
may be overestimating the binarity somewhat.} the stars are in binaries, so
the rate of supernova events in binaries is
$$
\alpha = 1 \times 10^{-2}~ {\rm{yr}}^{-1}~~.
\eqno(2.7)
$$

In the SN event, the resulting neutron star (NS) receives a kick.  For
the distribution of kicks, we take the measurements of  Cordes and Chernoff
(1997).  They found that the distribution is well represented by the sum of two
Gaussians with different standard dispersion $\sigma$
$$
\begin{array}{rcl}
80\%~ {\rm{have}}~ \sigma_1 & = & 175~ {\rm{km}}~ s^{-1} \\
20\%~ {\rm{have}}~ \sigma_2 & = & 700~ {\rm{km}}~ s^{-1}~~.
\end{array}
\eqno(2.8)
$$

If $v$ is the orbital velocity of the star which is becoming a
supernova, and $U$ is the kick given to the supernova remnant, a rough
approximation to the probability of the binary remaining together
is\footnote{Computer calculations with the program of Wettig and Brown
(1996) show this approximation to be accurate to $\leq$ 10\% in the region
of appreciable survival probabilities.  Survival with the Chernoff and Cordes
parameterization is slightly greater than that with the formula of Portegies
Zwart and Yungelson (1998).}
$$
\omega = \frac{ v^2}{U^2 + v^2}~~.
\eqno(2.9)
$$
\noindent If the kicks are distributed in a Gaussian, it is a fair
approximation to replace $U$ by $\sigma$, so
$$
\omega^\prime = \frac{v^2}{\sigma^2 + v^2}~~.
\eqno(2.10)
$$
\noindent Now for a binary of total mass $M=M_A + M_B$, in a circular
orbit of radius $a$, the orbital velocity is
$$
v^2 = GMa^{-1}~~.
\eqno(2.11)
$$
\noindent So if log $a$ is uniformly distributed, then log $v^2$ is
likewise, and
$$
d \phi = \frac{d(v^2)}{7v^2}~~,
\eqno(2.12)
$$
\noindent with limits on $v^2$ as indicated in the paragraph below
eq. (2.1), i.e.,
$$
30~ {\rm{km}} s^{-1} < v < 1000~ {\rm{km}} s^{-1}~~.
\eqno(2.13)
$$
\noindent Then the fraction of all binaries surviving the first
supernova event is
$$
{\cal{W}} = \int \omega^\prime d \phi = \frac{1}{7} \int \frac{d
(v^2)}{\sigma^2 + v^2} = \frac{1}{7} \ell n \frac{\sigma^2 +
v_{\rm{max}}^{~~~2}}{\sigma^2 + v_{\rm{min}}^{~~~2}}~~,
\eqno(2.14)
$$
\noindent where $v_{\rm{max}}$ and $v_{\rm{min}}$ are the limits
indicated in(2.13).  Taking into account the kick distribution (2.8),
$$
{\cal{W}} = \frac{0.8}{7} \ell n \frac{10^2 + 1.75^2}{0.3^2 + 1.75^2}
+ \frac{0.2}{7} \ell n  \frac{10^2 + 7^2}{0.3^2 + 7^2} = 0.43~~.
\eqno(2.15)
$$
\noindent Note that the minimum velocity, 30 km $s^{-1}$, is unimportant.
Changing this to 10 km $s^{-1}$ alters the result in (2.15) by only 10\%.
The maximum orbital velocity, i.e., the minimum orbital
radius, is relevant.  The high kick velocities, $\sigma_2$, contribute
only 0.03 of the 0.43 in eq. (2.15).  The rate of formation of binaries that
survive the first supernova event is now 0.43 per century.

A small correction should be made for the fact that the hydrogen envelope in
star $A$ does not generally extend out to the larger $a$ in our assumed
interval.  Weaver, Zimmerman \& Woosley (1978) find that for a $15\msun$ star
(star $A$) the envelope reaches out to $3.9 \times 10^8$ km in supergiant
stage.  Taking $M_B = 10\msun$, we find the Roche lobe to be $\sim$ 1/3 of the
distance between $A$ and $B$.  Thus, Roche Lobe overflow takes place only out
to $a \lsim 1.2 \times 10^9$ km in this example, less than our $4 \times 10^9$
km upper limit. However, relatively few systems at the higher $a$ survive the
kick velocities and we estimate the correction to be $\sim$ 10\%, which we
neglect.

It should also be noted that we have assumed the semi--major axis distribution
to be flat in log $a$ following mass transfer, whereas the empirical
determinations of it are before mass transfer.  From rough estimates this seems
to be a valid assumption.  We neglect all binaries which do not survive the
first phase of mass transfer and merge into single objects.  From Portegies
Zwart and Verbunt (1996), Table 4, this is about 20\%.  We have not made this
correction because in our calculations the close binaries do not survive
in--spiral.

\section{MASS TRANSFER}

The more massive star $A$ becomes a giant after time $t_A$.  Its Roche
lobe overflows and matter is transferred to star $B$.  This continues
until $A$ is stripped of its hydrogen envelope and thus reduced to a
He star.  In the mass range we are considering, $M_{AO}$ between 10
and perhaps 50 ${\rm M}_\odot$, the mass of the remaining He star is roughly
30\% of the initial mass.  Using a more accurate relation does not appreciably
change our results.  Denoting masses after this first transfer
by subscript 1,
$$
M_{A1} = 0.3 M_{AO}~~.
\eqno(3.1)
$$

Only part of the mass lost by $A$ will be attached to $B$.  We adopt
the estimate by Vrancken, DeGreve, Yungelson and Tutukov (1991)
(VDYT), that the fraction attached to $B$ is
$$
\beta \approx \left( M_{BO}/M_{AO} \right)^2 = q^2~~.
\eqno(3.2)
$$
\noindent (VDYT use the exponent 1.84).  Thus after transfer, the mass
of $B$ is
$$
M_{B1} = \left( q + 0.7 q^2 \right) M_{AO} \equiv M_{AO} f (q)~~.
\eqno(3.3)
$$
\noindent For $q=1$, the transfer is conservative, and $f(q) = 1.7$.
For $q < 0.68$, $f(q) < 1$.

We wish both stars to become supernovae at some stage in their life.
For star $A$, the condition is
$$
M_{AO} > M_{\sup} = 10 {\rm M}_\odot~~.
\eqno(3.4)
$$

It is here assumed that the minimum mass for a (type II or Ib)
supernova is $10 {\rm M}_\odot$.  For star $B$ to become a supernova, we
require that $M_{B1} > 10 {\rm M}_\odot$, in other words
$$
f (q) > 10 {\rm M}_\odot / M_{AO}~~.
\eqno(3.5)
$$

We  have assumed that $q$ is distributed uniformly between 0 and 1, so
the probability that star $B$ also becomes a SN, for any given
$M_{AO}$, is $1-q_a$, where
$$
f(q_a) = 10 {\rm M}_\odot/M_{AO}~~.
\eqno(3.6)
$$
Portegies  Zwart and Verbunt (1996) give in their Table 4 results for other
distributions in $q$.

The rate of supernovae in a given mass interval $d M_{AO}$ is given by
(2.5), hence the rate of birth of binaries that have supernovae in {\it
both} stars is
$$
\alpha^{\prime\prime} = \alpha n \int^{50 {\rm M}_\odot}_{10 {\rm M}_\odot} \left(
\frac{10{\rm M}_\odot}{M_{AO}} \right)^n \frac{d M_{AO}}{M_{AO}} \left[1-q_a
\left( M_{AO} \right) \right]~~.
\eqno(3.7)
$$
\noindent The upper limit has been taken to be $50\msun$ because above this mass
stars tend to lose their envelope mass in a luminous blue variable stage, and
become Wolf--Rayets.   They cannot then participate in common envelope
evolution. We have evaluated the integral numerically and find
$$
\alpha^{\prime\prime} = .50 ~\alpha~~.
\eqno(3.8)
$$
This value varies $\lsim$ 10\% with different reasonable Salpeter exponents
(Lattimer, 1997).

\section{EVOLUTION}

For masses near and above $10 {\rm M}_\odot$, the lifetime of a star in the
main sequence is about proportional to $M^{-2}$.  Thus, when star $A$
has come to the end of its main sequence life, after time $t_A$, star
$B$ will have gone through a fraction $q^2$ of its main sequence
lifetime.

The evolution time from the beginning of He burning to supernova is
roughly one-tenth of the main sequence evolution time.  This time is
relevant for us because mass will generally be transferred during helium core
burning, the supergiant state, when the hydrogen envelope is greatly extended.
During this time $0.1 t_A$, star $B$ has a mass $M_{AO} f (q)$, so it will go
through an added fraction $0.1 f^2 (q)$ of its main sequence life.  At the end
of this, the fraction of its main sequence life accomplished by star $B$ is
$$
g(q) = q^2 + 0.1 f^2 (q)~~.
\eqno(4.1)
$$
\noindent Table I gives the functions $f(q)$ and $g(q)$.  An
approximate formula is
$$
g(q) = 1.25 q^2 \pm 0.02~~.
\eqno(4.2)
$$

Of special interest is the value of $q$ for which $g(q) = 1$.
It is
$$
q_1 = 0.8897~~.
\eqno(4.3)
$$
\noindent The evolution is different according as $q < q_1$ or
$q>q_1$.   For $q < q_1$, we have Case I: Star $B$ is still in its main
sequence when star $A$ becomes a supernova.  The evolution in this
case will be treated in \S 5.  In Case II, $q>q_1$, star $B$ will be a
giant when star $A$ has its supernova event, this will be treated in
\S 6.

The probability, for any $M_{AO}$, that $q>q_1$ is
$$
1-q_1 = 0.110~~,
\eqno(4.4)
$$
\noindent so the rate of getting a case II binary is
$$
\alpha(II) = 0.11 \alpha~~,
\eqno(4.5)
$$
\noindent which is about 20\% of the total rate, (3.8).

\section{CASE I:  STAR B IN MAIN SEQUENCE}

\subsection{Possible Appearance}

As discussed in \S 4, Case I will hold if the initial mass ratio
$$
q = M_{BO}/M_{AO} < q_1 \approx 0.89~~.
\eqno(5.1)
$$
\noindent As shown after eq. (4.5), this applies to 80\% of all SN
binaries.  After star $A$ has had its supernova event, and provided
the binary stayed together (see \S 2), we have a neutron star $A$ and
a main sequence star $B$ of mass $M_{B1} = M_{AO} f (q)$ which must be
greater than 10${\rm M}_\odot$ because we want $B$ ultimately to become a
supernova.  So $B$ will be a $B$- or $O$-star.  The total lifetime of
such stars is of order $10^7$ years.  $A$ will be a neutron star, but
it will be a pulsar only for $5 \cdot 10^6$ years or less because its
spin will diminish by emitting pulsar radiation.  In fact, the binary
1259--63 with a $Be$ star companion and 0045--73 with a $B$ star companion
contain radio pulsars.  The former has a short 47 ms period and a double pulse
profile similar to the Crab; {\it i.e.}, it is a young pulsar. The latter
binary is in the small Magellanic cloud and is the most luminous binary radio
pulsar known.

Since the pulsar is unrecycled, the expected number of these binaries should be
compared with the single neutron star pulsar population, about 700 in number,
with more uncataloged.  This number should be multiplied by a factor $\sim 1/2$
for binarity and a factor of 0.43 (Eq. (2.15)) for breakup in the first
explosion.  This would leave the large number $\sim$ 150 if pulsars with
massive companions were as easily observable as single pulsars.  Of course,
pulsars are predominantly produced in the Galactic disc and, because of the
strong gravitational attraction of the dense matter in the disc, those with
massive companions will be unable to move out of the disc.  Stellar winds can
interfere with the radio pulses from these binaries, obscuring the narrower
ones.  Nonetheless, the factor necessary to reduce their observability is
startlingly large.  We return to this later.

\subsection{Structure of a Giant}

A giant has a He core, containing some 25\% of its mass, surrounded by
an envelope consisting mostly of H.  The envelope is usually in
convection so the entropy is constant.  The particles, nuclei and
electrons, are non-relativistic and thus have $\gamma = 5/3$,
therefore the envelope forms a polytrope of index $n = 3/2$.
Applegate (1997) has shown that the binding energy of the envelope is
$$
E_1 \cong (3/5) GM_B^{~~2} R^{-1}~~,
\eqno(5.2)
$$
\noindent where $R$ is the outer radius.  In this formula the
gravitational binding energy is decreased by 50\% by the kinetic energy, $E_1$
containing both effects.

After exhaustion of the core hydrogen, the radius $R$ increases first
slowly, then more rapidly, until it settles down (for masses in our
range) at several times $10^{13}$cm.

\subsection{Accretion}

Star $A$, beginning as a neutron star of mass 1.4 ${\rm M}_\odot$, sits at a
distance $r_A$ from the center of star $B$, awaiting the latter
to expand into a giant.  Its orbital period is a small fraction of the
time required for $B$ to evolve into a giant (which is of order $10^5$
years).  When $B$ reaches the radius $r_A$, material from $B$ begins
to accrete to $A$.  The rate of accretion is given by the
Bondi-Hoyle-Lyttleton theory,
$$
dM_A/dt = \pi \rho v R^2_{ac}~~,
\eqno(5.3)
$$
\noindent where $\rho$ is the density of the $B$-material, $v$ its
velocity relative to $A$, and $R_{ac}$ the accretion radius
$$
R_{ac} = 2 GM_A v^{-2}~~.
\eqno(5.4)
$$
\noindent Here $v$ is the velocity of the $B$ material relative to
$A$, which is essentially the orbital velocity of $A$ around $B$.
This is
$$
v^2 = G(M_B + M_A) a^{-1}~~.
\eqno(5.5)
$$
\noindent The energy loss is related to the accretion rate by (Iben \& Livio,
1993; Brown, 1995)
$$
\stackrel{\bullet}{E} = \frac{1}{2} c_d v^2 \stackrel{\bullet}{M}_A~~.
\eqno(5.6)
$$
\noindent Here $c_d$ is the drag coefficient (Shima, {\it et al.}, 1985),
$$
c_d = 2 \ell n \left( b_{\rm{max}}/b_{\rm{min}} \right) \approx 6~~,
\eqno(5.7)
$$
\noindent for highly supersonic flow.  (Iben and Livio, 1993, use $c_d = 8$.)
Here $b_{\rm{max}}$ and $b_{\rm{min}}$ are the maximum and
minimum impact parameters of the envelope material relative to the
neutron star.  Note that if $c_d$ decreases as the envelope moves towards
corotation with the neutron star, the $c_d$ in (5.6) must be replaced (Brown,
1995) by an effective coefficient of drag
$$
(c_d)_{eff} = c_d \left(\frac{{\cal{M}}^2 + 1}{{\cal{M}}^2}\right)^{3/2}
$$
where $\cal{M}$ is the Mach number $v/c_s$, with $c_s$ the speed of sound.  The
drop in $(c_d)_{eff}$ with decreasing Mach number is slower than in $c_d$.

The interaction energy of the two stars is
$$
E = \frac{1}{2} GM_A M_B a^{-1}~~,
\eqno(5.8)
$$
\noindent where $a$ is the semi-major axis of the orbit.  $(-E)$ is
the sum of potential and kinetic (orbital) energy.  The orbital
velocity is given by (5.5).  Generally, $M_A \ll M_B$, and we shall
neglect $M_A$ in (5.5).  Inserting (5.5) into (5.6),
$$
\stackrel{\bullet}{E} = \frac{1}{2} c_d GM_B a^{-1}
\stackrel{\bullet}{M}_A~~.
\eqno(5.9)
$$
\noindent Comparing (5.9) with (5.8), we see that we have two
variables depending on time, $M_A$ and
$$
Y = M_B a^{-1}~~.
\eqno(5.10)
$$
\noindent Taking the derivative of (5.8) and inserting in (5.9),
$$
M_A \stackrel{\bullet}{Y} + Y \stackrel{\bullet}{M}_A = c_d Y
\stackrel{\bullet}{M}_A~~.
\eqno(5.11)
$$
\noindent This can be integrated to give
$$
\ell n M_A = \left( c_d -1 \right)^{-1} \ell n Y +~ {\rm{const}}~~,
\eqno(5.12)
$$
$$
M_A~ \propto Y^{1/(c_d -1)} = Y^{1/5}~~.
\eqno(5.13)
$$
\noindent The final energy of the binary is then
$$
E_f = \frac{1}{2} GM_{A_i} Y_i \left( Y_f/Y_i \right)^{6/5}~~,
\eqno(5.14)
$$
\noindent where the subscript $i$ denotes the initial, $f$ the final
value of the respective quantities.

The binding energy $E_f$ of star $A$ to star $B$ serves to expel the
envelope of star $B$ whose initial binding energy is given by equation (5.2).
But it is believed that only a fraction of $E_f$ can serve to expel the
envelope, which is usually assumed to be one-half, hence
$$
E_f = 1.2 GM_{Bi}^{~~2} R^{-1}~~
\eqno(5.15)
$$
\noindent is necessary to completely eject the envelope.  In-spiral
begins when $R$ reaches $a_i$.  In-spiral is fast compared to the expansion of
the giant, therefore we set
$$
R = a_i~~.
\eqno(5.16)
$$
\noindent Then (5.15) becomes
$$
E_f = 1.2 GM_{Bi} Y_i~~.
\eqno(5.17)
$$
\noindent Inserting into (5.14) yields
$$
\left( Y_f/Y_i \right)^{1.2} = 2.4 M_{Bi}/M_{Ai}~~.
\eqno(5.18)
$$
\noindent Star $A$ is initially a neutron star, $M_{Ai} = 1.4$.  For
star $B$, we assume $M_{Bi} = 15$.  Then (5.18) yields
$$
Y_f/Y_i = 15~~.
\eqno(5.19)
$$

We use this first to find the result of accretion, with the help of
equation (5.13),
$$
M_{Af}/M_{Ai} = 1.73~~,
\eqno(5.20)
$$
$$
M_{Af} = 2.4\msun~~.
\eqno(5.21)
$$
\noindent Star $A$, by accretion, has become a black hole.  The lower limit for
a black hole, according to the argument of Brown and Bethe (1994), is $M_{min}
= 1.5 \msun$.  For the range $M_B = 10-20\msun$ the spread in $M_{Af}$ is
$2.25-2.5\msun$.  To within a few percent accuracy $M_{Af}/M_{Ai} =
(M_{Bi}/\msun)^{0.2}$, given the value we used for $c_d$.

Star $B$, by losing its envelope, becomes a He star.  We estimate
$$
M_{Bf}/M_{Bi} = 0.3~~.
\eqno(5.22)
$$
\noindent The size of the orbit is determined by equation (5.10),
$$
\frac{a_i}{a_f} = \frac{M_{Bi}}{M_{Bf}}~~ \frac{Y_f}{Y_i} = 50~~.
\eqno(5.23)
$$
\noindent The final distance between the stars, $a_f$, should not be
less than about $10^{11}$ so that the black hole $A$ is comfortably
outside the He star $B$.   On the other hand, if the two stars are to
merge within  a Hubble time, equation (7.9) shows that $a_f < 3.8
\times 10^{11}$.  So the initial distance of the two stars, after the
first mass exchange and the first supernova, should be
$$
0.5 \times 10^{13} < a_i < 1.9 \times 10^{13}~~{\rm{cm}}~~.
\eqno(5.24)
$$
\noindent The assumption is made here that the final black--hole, neutron--star
orbit has the same $a$ as the $a_f$ of the black--hole, He--star orbit.
Whereas the former will open out somewhat in the explosion of the He star, it
will generally be quite eccentric.  On the average this leads to an $\sim$ 40\%
increase in the $a_f$ for merger (Lattimer, 1997), so there will be
considerable cancellation between these neglected effects.  (See our later eq.
(7.10).)

If the initial distribution of distances is $da/7a$, the probability
of finding $a$ between the limits of equation (5.24), is
$$
P =  18\% \ .
\eqno(5.25)
$$
\noindent The probability of the binary surviving the first explosion was found
in (2.15) to be 0.43, so the combined probability is
$$
PW = 18 \times 0.43 = 7.7\% \ .
\eqno(5.26)
$$
\noindent Strictly speaking, the calculation of (2.15) should be done with the
orbital velocity limits corresponding to the limits of  $a_i$ in (5.24), which
are
$$
v_{max}^2 = 7.3 \times 10^{14} ({\rm cm~s}^{-1})^2
$$
$$
v_{min}^2 = 2.0 \times 10^{14} ({\rm cm~s}^{-1})^2 \ .
\eqno(5.27)
$$
\noindent The result is then
$$
PW = 8\% \ ,
\eqno(5.28)
$$
\noindent essentially unchanged.

The survival probability of the final He star explosion, in the
black--hole, He--star binary should be similar to that found by Wettig and
Brown (1996) for He--star, neutron--star binaries,
$$
P \simeq 50\% \ .
\eqno(5.29)
$$
\noindent An important ingredient in this relatively high probability is the
rather low mass of the He star before explosion.  As is well known, the binary
cannot stay together if its total mass after explosion is less than half that
before explosion.  The after--explosion mass is $2.4 + 1.4 = 3.8\msun$, so the
He star mass before explosion must be less than $0.2 \times 3.8 - 2.4 =
5.2\msun$.  Once the He stars have lost their H envelopes they rapidly lose
mass by wind, and those for the relevant ZAMS masses end up below $4\msun$.
For example, the initially $7\msun$ star of Woosley, Langer and Weaver (1995)
ends up at $3.2\msun$ before explosion.  Then if the SN kick is more or less
opposite to the orbital velocity before explosion, the binary will stay
together.

We now summarize the various factors entering into the black--hole, neutron
star evolution:
$$
R = 10^{-2} \times 0.50 \times 0.077 \times 0.5 = 1.9 \times 10^{-4}
{\rm  ~per~Galaxy~per~year} \ .
\eqno (5.30)
$$
\noindent These factors follow from eqs. (2.7), (3.8), (5.26), (5.29).
Because of our possible overestimates of binarity, of the number of systems
that undergo Roche Lobe overflow, and other effects mentioned earlier, this may
be somewhat too high.  We feel that a reasonable lower limit is
$$
R = 10^{-4} {\rm  ~per~Galaxy~per~year} \ .
\eqno(5.31)
$$
\noindent In fact, our rate per supernova ($\alpha^\prime/2 + \alpha +
\alpha^{\prime\prime}$) = 0.025 per year is 0.004 to be compared with 0.0036
from Portegies Zwart and Yungelson (1998) who have a supernova rate of 0.015
per year.  Thus, the chief difference between our result (5.31) and the $R =
5.3 \times 10^{-5}$ of these authors is due to the different assumed supernova
rate.  This shows that the many effects neglected in our calculations, but
included in their detailed computer calculations are unimportant for the
result.

Given our lower limit and the Wettig \& Brown (1996) $P = 50\%$, we would have
a rate of $2 \times 10^{-4}$ per Galaxy per year for formation of black--hole,
He--star binaries.  Given an average He--star lifetime of $5 \times 10^5$ yrs,
this would give $\sim$ 100 black--hole, He--star systems in the Galaxy.  This
is not far from the estimate of van den Heuvel (1995) who obtains 60
neutron--star, He--star systems, with a lower limit of $2.5\msun$ for the
He--star mass.  (With our lower limit of $2.2\msun$ for the He stars which end
up as neutron stars, his number would be $\gsim$ 50\% greater.)  Van den Heuvel
(1995) discusses reasons why only one such system, Cyg X--3, is seen.  If,
indeed, the neutron stars would not go into black holes, many of these neutron
stars would be recycled by the He--star wind, and this would lead to an
overproduction, by a large factor, of recycled binary pulsars.  We avoid this
overproduction problem by the neutron stars $A$ going into black holes.

Ergma \& Yungelson (1997) find by means of population synthesis the same
number 100 of black--hole, helium--star binaries in the Galaxy as do we.
However, their massive black holes arise directly from stars of ZAMS masses $>
30$ or $50\msun$, whereas our low mass black holes evolve from in spiral in the
hydrogen envelope of the compact object.  (We have not evolved systems with
massive black holes.)  Ergma and Yungelson suggest a combination of reasons
that, out of their many predicted systems, Cyg X--3 is the only example of such
systems seen.

\section{CASE II:  STAR B A GIANT}

\subsection{Evolution of Binary Neutron Stars}

Following hydrogen core burning, star $B$ will expand in a red giant phase,
which takes up $\geq$ 20\% of its lifetime.  In the first half of the red giant
phase the temperature in the center is not high enough to burn helium.  The
core undergoes contraction, raising its temperature. Because of the temperature
increase also just outside of the core, shell hydrogen burning begins and the
envelope expands modestly.  For a star of ZAMS mass $16\msun$ Bodenheimer and
Taam (1984) find that the radius increases out to $\sim 4 \times 10^{12}$ cm.
This first half of the red giant phase is not useful in this paper because if
the neutron star is met by the expanding giant envelope at $R < 0.5 \times
10^{13}$ cm, from condition (5.24) the neutron star will spiral into the core.
Interestingly, we find that this occurs for all of the standard high mass
X--ray binaries, such as SMC X--1, Cen X--3, LMC X--4, Vela X--1 and
4U1538--52.  The widest of these, Vela X--1, has $a = 0.37 \times 10^{13}$ cm.
Spiral--in for these objects has been found in many numerical calculations in
the literature.  In order to avoid spiral--in, we need binaries wider than the
HMXB's.

Relevant for us is the next $\sim$ 10\% of the lifetime of the star, the period
of helium core burning.  During this (supergiant) stage the star expands out to
several times $10^{13}$ cm.  If the first born neutron star is to escape
common--envelope evolution, then the two massive stars in the binary must burn
helium at the same time.  As we outline below, they can then expel their
hydrogen envelopes while burning helium.  The neutron star produced later then
has no hydrogen envelope.  As noted earlier, stellar evolutionary times go as
$M^{-2}$, so that stars $A$ and $B$ must be within 5\% of each other in mass,
if they are to burn helium at the same time.
However, if one star goes supernova,the other one will too.  Thus, we have a
rate of 0.05$\alpha$ (of eq. (2.7)) or $\sim$ 10\% of $\alpha^\prime$.

The fraction of these binaries with very nearly equal masses which survive
spiral in will be roughly equal to the fraction of black holes which survive in
our earlier discussion in \S 5, as we now argue.

If the two ZAMS masses are nearly equal, the two He stars tighten, expelling the
common H envelope (Brown 1995).  The binding energy of the H--envelope which
results from the two initial H envelopes is $\sim$ 4 times that of each
individual envelope (see eq. (5.2)).  If the He star were 4 times more massive
than the compact object discussed in \S 5 in common envelope evolution, the
final $a_f$ would be the same as in \S 5.  Since this is not far from being
true, the He stars will end up at roughly the same $a_f$ as found earlier for
the black hole and companion $B$.  In the explosion of the first He star,
Wettig and Brown (1996) found a survival probability of $\sim$ 50\%, not far
from the $\cal{W}$ of eq. (2.15).  The survival probability in the explosion of
the He star $B$ in the He--star, neutron--star binary will be similar to that in
the He--star, black--hole binary.  Thus, we can say that $\sim$ 10\% of the
original massive binaries will end up as binary pulsars.  This same ratio to
black--hole, neutron--star binaries holds for the probability of merging during
a Hubble time.  Given our rate of $10^{-4}$ per Galaxy per year for the latter
binaries, we find $10^{-5}$ per Galaxy per year for binary neutron stars.  This
may be an upper limit because He stars with ZAMS masses $\leq 15\msun$ expand
in the He shell burning phase and give the companion neutron star another
chance to go into a black hole (Brown 1997).  Using unpublished evolutionary
calculations of Woosley of He stars of mass $< 4\msun$ (roughly corresponding
to $15\msun$ ZAMS) in which wind mass loss is included, Fryer and Kalogera
(1997) show that only special conditions allow the pulsar in narrow
neutron--star binaries to avoid the envelope of these low--mass He stars.
Otherwise the pulsar accretes sufficient matter to go into a black hole, in
much the same way as was more crudely described by Brown (1997).  This implies
that most of the narrow binaries evolved from He star masses $< 4\msun$ in the
double He--star scenario are low--mass black--hole, neutron--star binaries.
Because of the much lower total mass in the He envelope, compared with the H
one (In Habets (1986), who does not include wind mass loss, the entire envelope
of a $2.5\msun$ He star above the helium burning shell is $\sim 0.8 \msun$.),
less mass will be accreted so that the black hole will be less massive than the
$\sim 2.4\msun$ of eq. (5.21).  This will decrease our estimate of $10^{-5}$
per Galaxy per year for merging binary neutron stars, probably by a factor
$\sim 2$.  This also adds a class, formed at about the same rate as binary
neutron stars, of low--mass black--hole, neutron star binaries in which the
black hole is not much more massive than the neutron star.
Again, our rate is in good
agreement with the $0.7 \times 10^{-5}$ found by Portegies Zwart and Yungelson
(1998) in their detailed computer evolution.

Note that our $10^{-5}$ per year per Galaxy is essentially the same as the
$\sim 8 \times 10^{-6}$ of van den Heuvel and Lorimer (1996).  These authors
increased their estimated number of binary pulsars which might be observed by a
beaming factor of 3 and a factor of 10 because 90\% are estimated to be
unobservable (Curran and Lorimer, 1995).  This factor of 30 is somewhat
uncertain, so it may be useful that we arrive at essentially the same rate
directly from an evolutionary calculation.

For very nearly equal initial masses ($q \sim$ 1) the mass transfer is
conservative.  In this case, from angular momentum transfer, we can deduce that
$$
\frac{a_f}{a_i} = \left(\frac{\mu_i}{\mu_f}\right)^2
\eqno(6.1)
$$
where
$$
\mu = M_AM_B/(M_A + M_B) \ .
\eqno(6.2)
$$
Since $\mu_f < \mu_i$, the final $\mu_f$ being slightly smaller than
$M_{Af}$, the He core mass in $A$, the orbits open out in conservative mass
transfer.  The main effect of this is to shift the logarithmic interval of
given binaries outwards.  But the magnitude of the logarithmic interval is
unchanged, so is (by our assumption) the number of binaries.

Note that if $q \sim$ 1 initially, the initial near equality of He core masses
of $A$ and $B$ will not be changed by the mass transfer.  Star $B$ will not be
rejuvenated by the transferred H--envelope from $A$ because there is not time
to cross the molecular weight barrier and convert the transferred H to He
(Braun and Langer, 1995).  Thus, the near equality in He--star masses in a
given binary is conserved.  This will lead to nearly equal neutron star masses
in a binary, except for a small correction from the accretion by the first
neutron star formed from the wind of its He--star companion (Brown,
1995).  This explains why the neutron star masses in a given binary tend to be
nearly equal. (In the standard scenario of evolution, where star $B$
is rejuvenated by mass transfer, the companion neutron star $B$ tends to be more
massive then the pulsar $A$.  This is the result found in the calculations of
Portegies Zwart (1997)).

\section{GRAVITATIONAL WAVES}

In \S 5 and 6 we have described how various types of compact binaries
are formed.  Once formed, they are subject to emission of gravitational
waves, as was demonstrated by Taylor and Hulse.  Shapiro and Teukolsky
(1983) discuss gravitational waves in their chapter~16,
and give a simple formula for the time required for a merger of the two
stars in their eq.~(16.4.10),
$$
T = \frac{5}{256}~~ \frac{c^5}{G^3 M^2 \mu} R^4~~.
\eqno(7.1)
$$
\noindent where $M= M_A + M_B$ and
$$
\mu = \frac{M_A M_B}{M_A + M_B}~~.
\eqno(7.2)
$$
\noindent The masses here are the masses after both stars have become
compact.

We are interested in the maximum initial distance permitted for the
two stars to merge in a Hubble time which we take to be
$$
T_H = 10^{10}~{\rm{yr}}~~.
\eqno(7.3)
$$
\noindent Then
$$
R_{\rm{max}}^{~~~~4} = 6.4 \frac{M_A M_B \left(M_A + M_B
\right)}{{\rm M}_\odot^{~~3}} R_{Sh}^{~~3} cT_H~~,
\eqno(7.4)
$$
\noindent where
$$
R_{Sh} = \frac{2G{\rm M}_\odot}{c^2} = 3.0~{\rm{km}}
\eqno(7.5)
$$
\noindent is the Schwarzschild radius of the sun.

Then
$$
R_{\rm{max}}^{~~~~4} = R_O^{~~4} \frac{M_A M_B \left(M_A + M_B
\right)}{{\rm M}_\odot^{~~3}}
\eqno(7.6)
$$
$$
R_O^{~~4} = 1.6 \times 10^{25} {\rm{km}}^4~~.
\eqno(7.7)
$$
\noindent Taking the masses of Case I, \S 5,
$$
M_A = 2.4 {\rm M}_\odot~~,~~ M_B = 1.4 {\rm M}_\odot
\eqno(7.8)
$$
\noindent we get
$$
R_{\rm{max}} = 3.8 \times 10^{11}~cm \approx 5{\rm R}_\odot~~.
\eqno(7.9)
$$
\noindent But we have shown in \S 5 that after expulsion of the
envelope, the distance between the two stars may be as low as $1
\times 10^{11} cm$ in which case the stars will merge in a small
fraction of the Hubble time.  The same is true of the Hulse-Taylor
binary, as is well known.

As noted following eq. (5.24), eccentricity in the final low--mass black--hole
neutron--star binary leads to an $a_f$ substantially larger than the $3.8
\times 10^{11}$ cm for merger of eq. (7.9).  In general most of the final
binaries will have $e > 0.5$, with a rapid rise just before $e = 1$.  The rise
results because preservation of the binary in the explosion is substantially
greater if the kick velocity is opposite to the orbital velocity before
explosion.  In this case the eccentricity $\epsilon$ is large.  the most
favorable situation is when the orbital and kick velocities are equal in
magnitude.  (See the figures in Wettig and Brown, 1996.)  Eggleton (1998) has
kindly furnished us with a useful interpolation formula for the increase.  The
factor to multiply the $T$ of eq. (7.1), which refers to circular orbits, is
$$
Z(e) \approx (1 - e^2)^{3.689 - 0.243\epsilon - 0.058e^2} ~~.
\eqno(7.10)
$$
\noindent This formula is accurate to about 1\% for $e \leq 0.99$.  Thus if the
initial eccentricity is 0.7 the time to shrink the orbit to zero is about 10\%
of the time required if the initial eccentricity were zero, for the same
initial period.  The maximum $a_f = 3.8 \times 10^{11}$ cm for circular orbits
would be increased by the fourth root of the decrease in time; i.e., up to $6.8
\times 10^{11}$ cm for this eccentricity.  The maximum $a_i$ in eq. (5.24)
would go up to $3.4 \times 10^{13}$ cm, increasing the favorable logarithmic
interval by $\sim$ 40\%.

We wish to carry out a complete  calculation of the distribution of
eccentricities using the computer program of Wettig and Brown (1996) before
assigning a definite number to the increase due to eccentricity.  Note that
this will affect our low--mass black--hole neutron star, neutron star binaries
in the same way, leaving the ratio nearly unchanged.  We note that Bloom,
Sigurdsson and Pols (1998), modifying the code created for binary evolution by
Pols and Eggleton (cf. Pols 1994 and references therein) find an extremely low
medium merging time of 2.1 million years for binary neutron stars.  This is
more than two orders of magnitude smaller than the merging times of the narrow
binaries 1913+16 and 1534+12, used to make estimates of the rate of mergers
from observation.  These binary neutron stars with very short lifetimes have
understandably not been observed, but should be added in estimates made from
observation, such as those of van den Heuvel and Lorimer (1996).

\section{``OBSERVABILITY PENALTIES" FOR BLACK--HOLE, NEUTRON--STAR
BINARIES}

With two supernovas per century, the formation rate of single pulsars is $\sim
1.25 \times 10^{-2}$ per year per Galaxy.  In fact, including binaries  as
noted following eq. (5.31) our total supernova rate is 0.025 per year.
However, in only $\sim 1/2$ of these cases is a neutron star formed, ZAMS
masses $\gsim 18\msun$ evolving into black holes according to the Brown and
Bethe (1994) scenario.  (We believe that SN(1987)A evolved into a low mass
black hole.)  The single pulsar formation rate of $\sim 1.25 \times
10^{-2}$ yr$^{-1}$ would seem to be the population with which
our black--hole, neutron--star binaries should be compared, since the neutron
star in both cases is unrecycled.  With a rate for the black--hole,
neutron--star binaries of $10^{-4}$ per year per Galaxy, one would see several
of these latter objects were this to be true.

As noted in \S 5a. the binarity in neutron--star, O, B--binaries seems to
severely inhibit observability, and it was suggested that these binaries
generally cannot get out of the Galactic disc where their radio signals will be
distorted by the dense matter.

The Hulse--Taylor pulsar is not far above the Galactic plane, at z = 0.26 kpc.
(Roughly half of the single pulsars are below this z, $\sim$ half above.)  The
Wolszczan 1534 is at high z = 0.51 kpc and the recently discovered 1518+49 is
at 0.57 kpc, both high.  Both of these are very old, with estimated ages of 250
and 16,000 Myr.  In all of these pulsars the magnetic field is low, $\sim
10^{10}$ G, so their spindown times are about two orders of magnitude greater
than those of fresh pulsars.  The unrecycled binary 2303+46 is at high z = 0.91
kpc, somewhat of a mystery.  According to our estimates of production rate of
$10^{-2}$ for single pulsars, and $10^{-5}$ for binary pulsars, with the latter
corrected by a factor $\sim 100$ for longer observability, one might expect to
see $\sim 70$ binaries, given $\sim 700$ single pulsars.  Curran and Lorimer
(1995), assuming the shape of the NS--NS luminosity function is similar to that
of normal pulsars, suggest that $\sim$ 90\% of these binary pulsars are missed
in current pulsar surveys.  This would remove most of the discrepancy noted
above.  Pulsars in the galactic plane are difficult to observe because the
plane is full of dust and electrons which absorb or scatter electromagnetic
radiation.  This applies to single as well as binary pulsars.  The difference
is, however, that pulsars in a binary have greater difficulty in getting out of
the galactic plane.  We suggest there is an ``observability penalty" which
results from binary pulsars moving more slowly out of the disc than single
neutron stars because of their higher mass.  This ``penalty" would be greater
for our black--hole, neutron--star systems, because they are more massive.

Furthermore, our black--hole, neutron--star binaries will generally be narrow,
with periods of $\sim 2-18$ hrs., more at the low end because of the $a^{-1}$
distribution of binaries.  Acceleration of the neutron star in its orbit in the
binary will make it harder to find the signal by Fourier analysis.

We have not been able to quantify these ``penalties", and offer the above as
only suggestions.

\section{CONCLUSION}

We believe that at least 1\% of massive binaries survive as binaries to
ultimate merger.  Taking the rate of supernovas in binaries to be one per
century per Galaxy, we find a merger rate of $10^{-4}$ per year per Galaxy, one
order of magnitude higher than believed before.

The most important assumption for this conclusion is that the semi--major axes
of binaries of heavy main--sequence stars are distributed as $da/a$, and that
this distribution extends out to $a = 2 \times 10^8$ km (even further once
eccentricity in the final binaries is taken into account), or orbital periods
as long as 100 days (eq. (5.24)).

We suggest that most of the mergings, leading to gravitational waves, will be
those of black--hole, neutron--star binaries, rather than binary neutron stars.
Earlier estimates are that the latter contribute $\sim 10^{-5}$ per year per
Galaxy.  In our evolutionary scenario the binary neutron star systems result
only from those cases in which the first born neutron star can escape the
common envelope evolution which otherwise sends it into a black hole.  This can
be realized if stars $A$ and $B$ are within $\sim$ 5\% of each other in mass.
Simple arguments show that this should result in $\sim$ 10\% of the binaries,
and scaling arguments show that about the same proportion of these selected
binaries survive spiral in and end up close enough to merge as in the
black--hole, neutron--star estimates.  The ten times higher rate for
black--hole, neutron--star mergers results, then, from the $\sim$ 90\% of
``failed" binary pulsar evolutions.  This ratio is robust, given our results
for the accretion onto the neutron star in the common envelope.

The reason for our high merger rate, $10^{-4}$, is that only a moderate
fraction of binaries get lost between the first supernova and the gravitational
merger.  In $\sim$ 50\% of the cases, the second star B is heavy enough to end
up in a supernova.  About 40\% survive the first supernova without splitting
the binary.  A small fraction is split by the second supernova kick.  And
another small fraction are lost because their separation, after both supernova
events, is too great to permit merger by gravitational wave emission within a
Hubble time.

A direct comparison with computer evolution is with case H of Portegies Zwart
and Yungelson (1998).  As noted following Eq. (5.31) and in our section on the
evolution of binary pulsars, both for the latter and for black--hole,
neutron--star binaries, agreement between our simple analytical evolution and
their detailed numerical calculations is good, $\lsim$ factor of 2 in the final
results.  With the same assumptions as to rate of supernovas they would be
closer and, most importantly, our ratio of black--hole, neutron--star mergings
to binary neutron--star mergings is nearly the same as theirs, 10.  This
indicates that the many detailed effects left out; {\it e.g.}, widening of the
orbits with mass exchange, are unimportant.  We believe that our analytical
work lends credence to the numerical work, and {\it vice versa}.  It is much
easier to assess changes that different effects would produce, so we believe
it worthwhile to have our simpler evolution.

Case H with inclusion of hypercritical accretion was only one of 8 cases
studied by Portegies Zwart and Yungelson (1998).  They remark that this case
fails to reproduce the short period binary pulsars.  However, inclusion of the
Wettig and Brown (1996) ``observability premium", which is now underway by
Portegies Zwart, should improve this.

In the standard scenario studied by Portegies Zwart and Yungelson, without
hypercritical accretion, masses of the pulsar progenitor come out substantially
less than those of the companion progenitor (Portegies Zwart, private
communication). This is easily understood by rejuvenation of the companion
following the initial mass transfer to it.  This is contrary to observation
where the pulsar is more massive than the companion neutron star in 1913+16 and
the two are nearly equal in mass in 1534+12.  The near equality in masses
follows from our double He star scenario, the somewhat greater pulsar mass in
1913+16 pinning down the progenitor mass (Brown, 1997).

A partial comparison can be made with Iben, Tutukov, and Yungelson (1995) who
used an extensive numerical scenario program developed at the Institute of
Astronomy in Moscow several years ago to carry out in much more detail than we
the evolution of binary compact objects. Iben, {\it et al.} obtain a rate of $3
\times 10^{-4}$ per year for merging neutron stars.  With inclusion of
hypercritical accretion in the common envelope evolution, 90\% of these would
become merging black--hole, neutron star binaries, to be compared with our rate
of $1.9 \times 10^{-4}$ per year (our eq. (5.30)).

Their birth rate for
HMXB's (our neutron star with massive companion) is $2.6 \times 10^{-3}$ per
year.  Our rate is $10^{-2} \times 0.58 \times 0.43 = 2.5 \times
10^{-3}$, with inclusion of kicks.  Their other results are difficult to
compare with ours, because of different basic assumptions.

Wettig and Brown (1996) suggested an ``observability premium" for the longer
spindown time of binary pulsars once their magnetic field had been brought down
(by accretion from the He--star wind during the He--star, neutron--star stage
in the Wettig and Brown evolution).  We suggest an ``observability penalty" for
the higher mass in our black--hole, neutron--star binaries.  This results from
their greater difficulty, due to their higher mass, to get out of the Galactic
disc.  In the latter, they are more difficult to observe.  Still, we believe
some of our black--hole, neutron--star binaries should be observable, and we
hope that our work is a challenge to observers.

Our results should be very important for LIGO. At the same time, LIGO becomes
very essential for testing our ideas about massive binaries:

Our rates in the Galaxy of $10^{-4}$ yr$^{-1}$ for\footnote{Here and in the
following BH means low--mass black hole.  We are presently preparing a paper
which estimates the rate for high--mass black--hole, neutron--star binaries
such as would result from Cyg X--1.} BH/NS binaries and $10^{-5}$
yr$^{-1}$ for NS/NS extrapolate to rates per unit volume in the universe of $8
\times 10^{-7}$Mpc$^{-3}$yr$^{-1}$ for BH/NS, and  $8 \times
10^{-8}$Mpc$^{-3}$yr$^{-1}$ for NS/NS (Phinney, 1991); which means that to see
one event per year, LIGO must look out to a distance of 70Mpc for BH/NS and
150Mpc for NS/NS.

Because of their larger masses, the BH/NS binaries can be seen farther by LIGO
than the NS/NS binaries.  Kip Thorne informs us that LIGO's first long
gravitational--wave search in 2002--2003 is expected to see BH/NS (assuming
masses of 2.4 and 1.4${\rm M}_\odot$) to a distance of about 35Mpc (too short
by a factor 2 according to our lower limit (5.31)), and NS/NS to 25Mpc (too
short by a factor 6).  However, enhancements of the initial LIGO
interferometers, planned for 2004, should reach out beyond 70Mpc for BH/NS,
bringing them into view; and other planned enhancements should reach NS/NS soon
thereafter; cf. Thorne (1998).

LIGO will measure each binary's chirp mass $M_{{\rm chirp}} = \mu^{3/5}M^{2/5}$
to an accuracy of a few tenths of a percent (Poisson and Will, 1995).  We
predict a bimodal distribution for these measurements:  The BH/NS systems
should display chirp masses concentrated near 1.7${\rm M}_\odot$; the NS/NS
systems should concentrate near 1.2${\rm M}_\odot$.  We predict a ratio of
event rates in LIGO of about 30 of the heavier systems for each of lighter one
(a factor 10 from rate per unit volume; a factor 3 from seeing BH/NS farther
than NS/NS).

Mergers of binary neutron stars and black--hole, neutron stars have been
considered in many papers as progenitors for gamma--ray bursters.  If one of
the neutron stars in the former does not collapse into a black hole, the
radiated neutrinos will deposit their energy in lifting baryons from the strong
gravitational potential instead of powering a relativistically expanding
pair--plasma fireball.  This can be avoided if one of the compact objects is a
black hole, in which case baryons drop through the event horizon carrying their
binding energy with them (Ruffert and Janka, 1998).  This leaves a low baryon
contamination of the fireball, which cannot be larger than $\sim 10^{-5}
\msun$.  The black hole must be of mass less than $5\msun$; otherwise
the neutron star would be swallowed before disruption (Rees, 1997).  Our many
low--mass black--hole, neutron--star binaries offer a heretofore unexpected
number of possible progenitors.

\acknowledgments

The authors would like to thank Simon F. Portegies Zwart for many useful
discussions and for providing them with unpublished material.  We thank Lev
Yungelson for correspondence and for sending us the Ergma, Yungelson paper in
advance of publication.  We are also grateful to C.-H. Lee for using the Wettig
and Brown (1996) code to check a number of analytical calculations.  We would
like to thank Zaven Arzoumanian and Steve Thorsett for much advice about
observability, and David Chernoff for giving us the results of Cordes and
Chernoff (1997) long before publication. One of us (G.E.B.) is grateful to
Madappa Prakash for many helpful conversations.  Jim Lattimer checked many of
our calculations and gave many helpful criticisms.  G.E.B. was partially
supported by the U.S. Department of Energy under grant No. DE-FG02-88ER40388.
Kip Thorne helped us to relate our results more quantitatively to LIGO.  We are
grateful to Peter Eggleton for advice on the distribution of massive binaries
and for the interpolation formula (7.10) for the effect of eccentricity.

\newpage
\centerline{ REFERENCES}

\noindent Applegate, J. H. 1997 Columbia Univ. Preprint\\
\noindent Bloom, J. S., Sigurdsson, S., \& Pols, O. R. 1998, MNRAS, to be
published \\
\noindent Bodenheimer, P., \& Taam, R. E. 1984, ApJ, 280, 771 \\
\noindent Braun, H., \& Langer, N. 1995, A\&A, 297, 771\\
\noindent Brown, G. E. 1995, ApJ, 440, 270\\
\noindent Brown, G. E. 1997, Phys. Bl., 53,  671\\
\noindent Brown, G. E., \& Bethe, H. A. 1994, ApJ, 423, 659 \\
\noindent Cordes, J. M., \& Chernoff, D. F. 1997 Cornell Preprint\\
\noindent Curran, S. J., \& Lorimer, D. R. 1995, MNRAS, 276, 347\\
\noindent Eggleton, P. 1998, private communication \\
\noindent Ergma, E., \& Yungelson, L. R. 1997, A\&A, to be published \\
\noindent Fryer, C., \& Kalogera, V. 1997, ApJ, 489, 244 \\
\noindent Garmany, C. D., Conti, P. S., \& Massey, P. 1980, apJ, 242, 1063 \\
\noindent Habets, G. M. H. J. 1986, A\&A, 165, 95 \\
\noindent Iben, I., \& Livio, M. 1993, PASP 105, 1373 \\
\noindent Iben, I., Tutukov, A. V., \& Yungelson, L. R. 1995, ApJSS, 55, 100, 217 \\
\noindent Lattimer, J. M. 1997, private communication \\
\noindent Phinney, E. S. 1991, ApJL, 380, L17 \\
\noindent Poisson, E., \& Will, C. M. 1995, Phys. Rev. D, 52, 848 \\
\noindent Pols, O. R. 1994, A\&A, 290, 119 \\
\noindent Portegies Zwart, S. F. 1997, private communication \\
\noindent Portegies Zwart, S. F. \& Verbunt, F. 1995, A\&A, 309, 179 \\
\noindent Portegies Zwart, S. F. \& Yungelson, L. R. 1998, A\&A, to be
published \\
\noindent Rees, M. J. 1997, astro--phy/9701162 \\
\noindent Ruffert, M., \& Janka, H.-J. 1998, Proc. 4th Huntsville Gamma--Ray
Burst Symposium, \\ \indent Huntsville, Alabama, Sept. 15--20, 1997. AIP Conf.
Proc., eds. Meegan, C. A., \\ \indent \& Cushman, P., AIP, New York \\
\noindent Shapiro, S. L., \& Teukolsky, S. A. 1983, {\it Black Holes, White
Dwarfs and Neutron Stars}, \\ \indent John Wiley \& Sons, New York\\
\noindent Shima, E., Matsuda, T., Takeda, H., \& Sawada, K. 1985, MNRAS, 217,
367 \\
\noindent Thorne, K. S. 1998, in {\it Black Holes and Relativistic Stars}, ed.
Wald, R. M., University of \\ \indent Chicago Press, Chicago \\
\noindent Van den Heuvel, E. P. J. 1995, J. Astrophys. Astr., 16, 255 \\
\noindent Van den Heuvel, E. P. J., \& Lorimer, D. R. 1996, MNRAS, 283, L37\\
\noindent Vrancken, J., DeGreve, J. P., Yungelson, L., \& Tutukov, A. 1991,
A\&A, 249, 411\\
\noindent Weaver, T. A., Zimmerman, G. B., \& Woosley, S. E. 1978, ApJ, 225,
1021 \\
\noindent Wettig, T., \& Brown, G. E. 1996 New Astron., 1, 17 \\
\noindent Woosley, S. E., Langer, N., \& Weaver, T. A. 1995, ApJ, 448, 315 \\

\newpage
\centerline{TABLE CAPTIONS}

\noindent TABLE 1. NOTE.--  Mass of star $B$ after mass transfer, $f(q)
= M_{B1}/M_{AO}$.  Fraction $g(q)$ of main sequence accomplished by
$B$ when $A$ has
finished its MS.  $1.25 q^2$ is shown for comparison
with $g(q)$.

\newpage

\begin{table*}[t]
\begin{center}
\centerline {TABLE~1}
\vskip 8pt
\begin{tabular}{llll}
\hline\hline
$q$ & $f(q)$ & $g(q)$ & $1.25 q^2$ \\
\hline
0.95 & 1.582 & 1.153 & 1.128  \\
0.9 & 1.467 & 1.025 & 1.012 \\
0.89 & 1.444 & 1.001 & 0.990  \\
0.85 & 1.356 & 0.906 & 0.903  \\
0.8 & 1.248 & 0.796 & 0.800 \\
0.7 & 1.043 & 0.599 & 0.613  \\
0.6 & 0.852 & 0.433 & 0.450 \\
0.5 & 0.675 & 0.296 & 0.313 \\
0.3 & 0.363 & 0.103 & 0.112 \\
0.1 & 0.107 & 0.011 & 0.013 \\
\hline\hline
\end{tabular}
\end{center}
\end{table*}

\clearpage

\end{document}